\documentstyle[12pt]{article}

\def\thebibliography#1{\section*{
References}\list
  {\arabic{enumi}.}{\settowidth\labelwidth{#1}\leftmargin\labelwidth
    \advance\leftmargin\labelsep
    \usecounter{enumi}}
    \def\newblock{\hskip .11em plus .33em minus .07em}
    \sloppy\clubpenalty4000\widowpenalty4000
    \sfcode`\.=1000\relax}

\def\op#1{\mathop{\fam0 #1}\limits}

\newcommand{\pr}{{\rm pr}}

\newcommand{\Ker}{{\rm Ker\,}}
\newcommand{\im}{{\rm Im\, }}
\newcommand{\nm}[1]{\mid {#1}\mid}

\newcommand{\beq}{\begin{equation}}
\newcommand{\eeq}{\end{equation}}
\newcommand{\ben}{\begin{eqnarray}}
\newcommand{\een}{\end{eqnarray}}
\newcommand{\be}{\begin{eqnarray*}}
\newcommand{\ee}{\end{eqnarray*}}
\newcommand{\bea}{\begin{eqalph}}
\newcommand{\eea}{\end{eqalph}}

\newcommand{\cA}{{\cal A}}

\newcommand{\rA}{{\rm Ann\,}}

\newcommand{\cT}{{\cal T}}
\newcommand{\cP}{{\cal P}}
\newcommand{\cL}{{\cal L}}
\newcommand{\cV}{{\cal V}}
\newcommand{\cE}{{\cal E}}
\newcommand{\cN}{{\cal N}}
\newcommand{\cR}{{\cal R}}
\newcommand{\cH}{{\cal H}}
\newcommand{\cF}{{\cal F}}

\newcommand{\cS}{{\cal S}}

\newcommand{\bL}{{\bf L}}
\newcommand{\bQ}{{\bf Q}}
\newcommand{\bR}{{\bf R}}
\newcommand{\bC}{{\bf C}}

\newcommand{\dl}{\delta}
\newcommand{\la}{\lambda}

\newcommand{\f}{\phi}

\newcommand{\om}{\omega}
\newcommand{\Om}{\Omega}
\newcommand{\m}{\mu}

\newcommand{\g}{\gamma}
\newcommand{\G}{\Gamma}

\newcommand{\th}{\theta}

\newcommand{\up}{\upsilon}
\newcommand{\vt}{\vartheta}

\newcommand{\si}{\sigma}

\newcommand{\z}{\zeta}

\newcommand{\sh}{\sharp}

\newcommand{\bom}{{\bf\Omega}}
\newcommand{\bth}{{\bf\Theta}}

\newcommand{\w}{\wedge}
\newcommand{\wt}{\widetilde}
\newcommand{\wh}{\widehat}
\newcommand{\ol}{\overline}
\newcommand{\dr}{\partial}
\newcommand{\lla}{\op\longleftarrow}
\newcommand{\ar}{\op\longrightarrow}

\newcommand{\llra}{\longleftrightarrow}
\newcommand{\ot}{\otimes}
\newcommand{\ap}{\approx}

\newcommand{\rrq}{{\ol q}}

\let\ssection=\section
\renewcommand{\section}{\setcounter{equation}{0}\ssection}

\newcounter{eqalph}[section]
\newcounter{equationa}[section]
\newcounter{example}[section]
\newcounter{remark}[section]
\newcounter{theorem}[section]
\newcounter{proposition}[section]
\newcounter{lemma}[section]
\newcounter{corollary}[section]
\newcounter{definition}[section]

\def\theremark{\arabic{section}.\arabic{remark}}

\def\thedefinition{\arabic{section}.\arabic{definition}}

\newenvironment{proof}{\noindent {\it Proof.}}{$\Box$
\medskip }

\newenvironment{rem}{\refstepcounter{remark} \medskip\noindent{\it Remark
\theremark.}}{ \medskip }
\newenvironment{theo}{\refstepcounter{definition} \medskip\noindent{\bf
Theorem \thedefinition}.\it}{\medskip }
\newenvironment{prop}{\refstepcounter{definition} \medskip\noindent{\bf
Proposition \thedefinition}.\it}{\medskip }
\newenvironment{lem}{\refstepcounter{definition} \medskip\noindent{\bf Lemma
\thedefinition}.\it }{\medskip }
\newenvironment{cor}{\refstepcounter{definition} \medskip\noindent{\bf
Corollary \thedefinition}.\it }{\medskip }

\newenvironment{eqalph}{\stepcounter{equation}
\setcounter{equationa}{\value{equation}}
\setcounter{equation}{0}

\begin{eqnarray}}{\end{eqnarray}
\setcounter{equation}{\value{equationa}}}

\newcommand{\nw}[1]{[{#1}]}
\newcommand{\der}{\rm Der}

\begin{document}
\hbox{}

\noindent
{\large \bf Constraints in  Hamiltonian time-dependent mechanics}
\newline\newline
{\sc Giovanni Giachetta$^1$, Luigi Mangiarotti$^1$ \newline and Gennadi
Sardanashvily$^2$}
\newline{\small
$^1$ Department of Mathematics and Physics, Camerino University, 62032
Camerino, Italy\newline
E-mail: mangiaro@camserv.unicam.it\newline
$^2$ Department of Theoretical Physics, Moscow State University, 117234
Moscow, Russia} \newline
E-mail: sard@grav.phys.msu.su\newline\newline
{\bf Abstract} 
In Hamiltonian time-dependent mechanics, the  
Poisson bracket does not define dynamic equations, that   implies the
corresponding peculiarities of describing time-dependent holonomic
constraints. As in conservative mechanics, one can consider the Poisson
bracket of constraints, separate them in first and second class constraints,
construct the Koszul--Tate resolution and a BRST complex. However, the Poisson
bracket of constraints and a Hamiltonian makes no sense.
Hamiltonian vector fields for first class constraints are not
generators of gauge transformations. In the case of Lagrangian constraints, we
state the comprehensive relations between solutions of the Lagrange equations
for an almost regular Lagrangian and solutions of the Hamilton equations for
associated Hamiltonian forms, which live in the Lagrangian constraint space.
Degenerate quadratic Lagrangian systems are studied in details.  We construct
the Koszul--Tate resolution for Lagrangian constraints of these systems in an
explicit form.

\tableofcontents

\section{Introduction}

The
technique of Poisson and symplectic manifolds is well known to provide the
adequate Hamiltonian formulation of classical and quantum conservative
mechanics. This is also the case of presymplectic Hamiltonian
systems. Since every presymplectic form can be represented as a pull-back of a
symplectic form by a coisotropic imbedding \cite{got82,book98},
 a presymplectic Hamiltonian system can be seen as a Dirac
constraint system \cite{cari85,book98}. An autonomous Lagrangian system
also exemplifies a presymplectic Hamiltonian system where a
presymplectic form is the exterior differential of the Poincar\'e--Cartan
form, while a Hamiltonian is the energy function
\cite{cari93a,leon96,book98,mun}.
A generic example of conservative Hamiltonian mechanics 
is a regular Poisson manifold $(Z,w)$ where a Hamiltonian is a real
function $\cH$ on $Z$. Given the corresponding Hamiltonian vector field 
$\vt_\cH=w^\sh(df)$, the closed subbundle $\vt_\cH(Z)$ of the tangent
bundle
$TZ$ is an 
autonomous first order dynamic equation on a manifold $Z$, called the Hamilton
equations. The evolution equation on the Poisson algebra $C^\infty(Z)$ is the
Lie derivative $ \bL_{\vt_\cH}f= \{\cH,f\}$,
expressed into the Poisson bracket of the Hamiltonian $\cH$ and 
functions $f$ on $Z$. This
description, however, cannot be extended in a straightforward manner to
time-dependent mechanics subject to time-dependent transformations.

The existent formulations of time-dependent mechanics
 imply usually a preliminary splitting
of a configuration space $Q=\bR\times M$ and a momentum phase space  
$\Pi=\bR\times Z$,
where $Z$ is a Poisson manifold \cite{cari93,chinea,eche,ham,mora,leon93}.
From the physical viewpoint, this means that a certain reference frame is
chosen. In this case, the momentum phase space $\Pi$ is endowed with the
Poisson product of the zero Poisson structure on
$\bR$ and the Poisson structure on $Z$. A
Hamiltonian is defined as a real function $\cH$ on
$\Pi$. The corresponding Hamiltonian vector field $\vt_\cH$ on 
$\Pi$ is vertical with respect to
the fibration $\Pi\to\bR$. Due to the canonical imbedding 
\beq
\Pi\op\times_\bR T\bR\to T\Pi, \label{mm6}
\eeq
one
introduces the vector field
\beq
\g_\cH=\dr_t +\vt_\cH, \label{mm0}
\eeq
where $\dr_t$ is the standard vector field on $\bR$ \cite{ham}. The first order
dynamic equation $\g_\cH(\Pi)\subset T\Pi$ on the manifold $\Pi$ plays the
role of Hamilton equations. The evolution equation on the Poisson algebra
$C^\infty(\Pi)$ is given by the Lie derivative
\beq
\bL_{\g_\cH}f= \dr_t f +\{\cH,f\}. \label{mm7}
\eeq

This is not the case of mechanical systems subject to time-dependent
transformations. These transformations, including canonical and
inertial frame transformations, violate the splitting
$\bR\times Z$. As a consequence, there is no
canonical imbedding (\ref{mm6}), and the vector field (\ref{mm0}) is not well
defined. At the same time, one can treat the imbedding (\ref{mm6}) as a
trivial connection on the bundle $\Pi\to\bR$, while $\g_\cH$
(\ref{mm0}) is the sum of the horizontal lift onto $\Pi$ of the vector field
$\dr_t$ by this connection and of the vertical vector field
$\vt_\cH$. This observation make us to think of non-relativistic
time-dependent mechanics as being a particular field theory on fibre bundles
over $\bR$, where the time axis $\bR$ is parameterized by the
Cartesian coordinates
$t$ with the transition functions 
$t'=t+$const. Then $\bR$ is
provided with the above mentioned standard vector field $\dr_t$ and the
standard 1-form
$dt$. Every fibre bundle over $\bR$ is obviously trivial, but
its trivialization is not necessarily canonical.

\begin{rem} 
The following peculiarity of bundles over $\bR$ is important. Let
$Y\to\bR$ be a fibre bundle coordinated by $(t,y^A)$, and
$J^1Y$ its first order jet manifold, equipped with the adapted coordinates
$(t,y^A,y^A_t)$. There is the canonical imbedding 
\beq
\la=\dr_t+y^A_t\dr_A: J^1Y\op\hookrightarrow_Y TY \label{mm8}
\eeq
onto the affine subbundle 
of $TY\to Y$ of elements $\up\in TY$ such that
$\up\rfloor dt=1$. This subbundle is modelled over the vertical tangent bundle
$VY\to Y$. As a consequence, there is one-to-one correspondence between the
connections $\G$ on the fibre bundle $Y\to \bR$, treated as sections of the
affine jet bundle $\pi^1_0:J^1Y\to Y$ \cite{book99}, and the nowhere vanishing
vector fields $\G=\dr_t +\G^A\dr_A$ on $Y$, called 
horizontal vector fields,
 such that $\G\rfloor dt=1$ \cite{book98,book99}. The corresponding
covariant differential reads
\be
D_\G=\la-\G:J^1Y\op\to_Y VY, \qquad \dot y^A\circ D_\G=y^A_t-\G^A.
\ee
Let us also recall  the total derivative 
$d_t =
\dr_t +y^A_t\dr_A +\cdots$ and the exterior algebra
homomorphism
\beq
h_0:\f dt+ \f_A dy^A \mapsto (\f+\f_A y^A_t) dt \label{cmp100}
\eeq
which sends exterior forms on $Y\to\bR$ onto the horizontal forms on
$J^1Y\to \bR$,
and vanishes on contact forms $\th^A=dy^A
-y^A_tdt$. 
\end{rem}

Lagrangian time-dependent mechanics follows
directly Lagrangian field theory
\cite{giach92,krupkova,leon97a,book98,massa}. It implies the
existence of a configuration space $Q\to\bR$ of a mechanical system, 
and a Lagrangian is
defined as a horizontal density 
\beq
L=\cL dt, \qquad \cL: J^1Q\to\bR, \label{mm43}
\eeq
on the velocity phase space $J^1Q$.
However, there is the essential difference 
between field
theory and time-dependent mechanics. 
The curvature of any connection $\G$ on 
a configuration bundle $Q\to\bR$ vanishes
identically, and these
connections fail to be dynamic variables, but characterize reference frames.
 The horizontal vector
field $\G$ sets a tangent vector at
each point of the configuration space
$Q$, which can be seen as the velocity of an "observer" at this point
\cite{book98,massa,sard98}. There is the correspondence
between the connections $\G$ on the configuration bundle $Q\to\bR$ and the
trivializations of $Q\to\bR$ such that $\G=\dr_t$ in the adapted coordinates
(see Section 2). 

A generic momentum phase space of time-dependent mechanics
is a fibre bundle $\Pi\to \bR$ endowed with a regular Poisson structure whose
characteristic distribution belongs to the vertical tangent bundle $V\Pi$ of 
$\Pi\to \bR$ \cite{ham}. Such a Poisson structure however cannot provide 
dynamic equations. A first order dynamic equation on  
$\Pi\to\bR$, by definition, is a section of the affine jet bundle $J^1\Pi\to
\Pi$, i.e., a connection on $\Pi\to\bR$. Being a horizontal vector
field, such a
connection  cannot be a Hamiltonian vector field with respect to the
above mentioned Poisson structure on $\Pi$. 

One can overcome this difficulty as follows.
Let $Q\to\bR$ be a configuration bundle of time-dependent mechanics.
The corresponding momentum phase space is the vertical cotangent bundle
$\Pi=V^*Q\to \bR$, called the Legendre bundle, while the cotangent bundle
$T^*Q$ is the homogeneous momentum phase
space. $T^*Q$ admits the canonical Liouville form
$\Xi$ and the symplectic form
$d\Xi$, together with the  corresponding non-degenerate Poisson
bracket $\{,\}_T$ on the ring $C^\infty(T^*Q)$.
Let us consider the subring of $C^\infty(T^*Q)$ which comprises the
pull-backs $\zeta^*f$ onto $T^*Q$ of functions $f$ on the vertical
cotangent bundle $V^*Q$ by the canonical fibration 
\beq
\zeta:T^*Q\to V^*Q. \label{mm5}
\eeq
This subring is closed under the Poisson bracket
$\{,\}_T$, and
$V^*Q$ is provided with the regular Poisson structure $\{,\}_V$ such that 
\beq
\zeta^*\{f,g\}_V=\{\zeta^*f,\zeta^*g\}_T\label{m72'}
\eeq
\cite{vais}. Its characteristic
distribution coincides with the
vertical tangent bundle $VV^*Q$ of $V^*Q\to \bR$. 
Given  a section $h$ of the  bundle (\ref{mm5}), let us consider the
pull-back forms
\beq
\bth =h^*(\Xi\w dt), \qquad \bom=h^*(d\Xi\w dt) \label{z401}
\eeq
on $V^*Q$, but these forms are independent of a section
$h$ and are canonical exterior forms on $V^*Q$. The pull-backs $h^*\Xi$ are
called the Hamiltonian forms.
 With $\bom$, 
the Hamiltonian vector field  
$\vt_f$ for a function $f$ on $V^*Q$ is given by the relation
\beq
\vt_f\rfloor\bom = -df\w dt, \label{z406}
\eeq
while the Poisson bracket (\ref{m72'}) is written as
\be
\{f,g\}_Vdt=\vt_g\rfloor\vt_f\rfloor\bom.
\ee 
Note that a generic momentum
phase space
$\Pi\to\bR$ of time-dependent mechanics can be seen locally as
the Poisson product over $\bR$ of the Legendre bundle $V^*Q\to\bR$ and
a fibre bundle over
$\bR$, equipped with the zero Poisson structure.

The pair $(V^*Q,\bom)$ is the particular
$(n=1)$-polysymplectic phase space of the covariant Hamiltonian field theory
(see 
\cite{cari,book,got91,sard95} for a survey). Following its general scheme, we
can formulate the Hamiltonian time-dependent mechanics as follows
\cite{book98,sard98}.

A connection $\g$ on the Legendre bundle
$V^*Q\to\bR$ is called canonical if the corresponding horizontal vector
field is canonical for the Poisson structure on $V^*Q$, i.e., the form
$\g\rfloor\bom$ is closed. We will prove that such a form is necessarily exact.
A canonical connection $\g$ is a said to be a Hamiltonian connection if 
\beq
\g\rfloor\bom= dH \label{z405}
\eeq
where $H$ is a Hamiltonian form on $V^*Q$. We show that every Hamiltonian form
admits a unique Hamiltonian connection $\g_H$, and that any canonical
connection is locally a Hamiltonian one. Given a Hamiltonian form $H$, 
the kernel of the covariant differential $D_{\g_H}$, associated with the
Hamiltonian connection $\g_H$, is a closed imbedded subbundle of the jet
bundle
$J^1V^*Q\to \bR$, and so is the system of first order PDEs on 
the Legendre bundle 
$V^*Q\to \bR$. These are 
the Hamilton equations in time-dependent mechanics, while the Lie derivative 
\beq
\bL_{\g_H}f=\g_H\rfloor df
\label{m59}
\eeq
defines the evolution equation on $C^\infty(V^*Z)$. As in
the polysymplectic case \cite{book,sard94,sard95},
this Hamiltonian dynamics  is equivalent  to the Lagrangian one for
hyperregular Lagrangians, while a degenerate
Lagrangian involves a set of associated Hamiltonian forms in
order to exhaust solutions of the Lagrange equations.

The main peculiarity of Hamiltonian time-dependent mechanics
lies in the fact that, since $\g_H$ is not a
vertical vector field, the right-hand side of the evolution equation
(\ref{m59}) is not expressed into the Poisson bracket in a canonical way, but
contains a frame-dependent term.
Every connection $\G$ on the
configuration bundle $Q\to\bR$ is an affine section of
the  bundle (\ref{mm5}), and defines the Hamiltonian form $H_\G=\G^*\Xi$ on
$V^*Q$. The corresponding Hamiltonian connection is the canonical lift 
$V^*\G$ of $\G$ onto the Legendre bundle $V^*Q$ \cite{book,book99}.
Then  any Hamiltonian form $H$ on $V^*Q$ admits splittings
\ben
&& H=H_\G -\wt\cH_\G dt, \label{m46'}\\
&& \g_H=V^*\G + \vt_{\wt\cH_\G}, \nonumber
\een
where $\vt_{\wt\cH_\G}$ is the vertical Hamiltonian field for the function
$\wt\cH_\G$, which the energy
function with respect to the reference frame $\G$ (see Section 4). With the
splitting (\ref{m46'}), the evolution equation (\ref{m59}) takes the form
\beq
\bL_{\g_H}f= V^*\G\rfloor H +\{\wt\cH_\G,f\}_V. \label{m96}
\eeq

Let the configuration bundle $Q\to\bR$ with an $m$-dimensional typical
fibre $M$ be coordinated by
$(t,q^i)$. Then Legendre bundle $V^*Q$ and the cotangent bundle
$T^*Q$ are provided with holonomic coordinates $(t,q^i,p_i=\dot q_i)$ and
$(t,q^i,p_i,p)$, respectively. Relative to these
coordinates, a Hamiltonian form
$H$ on $V^*Q$ reads
\beq
H=h^*\Xi= p_i dq^i -\cH dt. \label{b4210}
\eeq
It is the well-known integral
invariant of Poincar\'e--Cartan, where $\cH$ is a Hamiltonian
in time-dependent mechanics. A glance at the expression (\ref{b4210}) shows
that $\cH$ fails to be a scalar under time-dependent
transformations. 
Accordingly, the evolution equation (\ref{m96}) takes the local form
\beq
\bL_{\g_H}=\dr_tf +\{\cH,f\}_V, \label{mm15}
\eeq
but one should bear in mind that the terms in its right-hand side, 
taken separately, are not well-behaved objects under time-dependent
transformations. In particular, the equality
$\{\cH,f\}_V=0$ is not preserved under time-dependent
transformations.

The above peculiarities of Hamiltonian time-dependent mechanics 
imply the corresponding peculiarities of describing time-dependent holonomic
constraints. As in conservative mechanics, one can consider the Poisson
bracket of constraints, separate them in first and second class constraints,
construct the Koszul--Tate resolution and BRST complex. However, the Poisson
bracket of constraints and a Hamiltonian makes no sense in time-dependent
mechanics. Hamiltonian vector fields for first class constraint functions are
not generators of gauge transformations. We will pay a special attention to
Lagrangian constraints.  Every Lagrangian
$L$ defines the Legendre map
\beq
\wh L:J^1Q \op\to_Q V^*Q, \qquad p_i\circ\wh L =\pi_i,\label{a303}
\eeq 
whose image $N_L=\wh L(J^1Q)\subset V^*Q$ is called the Lagrangian constraint
space. We state the comprehensive relationship between solutions of the
Lagrange equations for an almost regular Lagrangian $L$ and solutions in $N_L$
of the Hamilton equations for associated Hamiltonian forms. The detailed
analysis of degenerate quadratic Lagrangian systems in Section 7 is
appropriate for application to many physical models. In Section 9, we
construct the Koszul--Tate resolution for Lagrangian constraints of such a
degenerate system in an explicit form.

\section{Interlude I. Non-relativistic reference frames}

As was mentioned above, a reference frame in non-relativistic mechanics is
identified with a connection 
$\G$ on the configuration bundle
$Q\to\bR$. Being flat, every 
connection $\G$ on $Q\to\bR$ 
yields an integrable 
horizontal distribution on $Q$, whose integral manifolds 
are integral curves  of the horizontal vector field $\G$ which are
transversal to the fibres of the  bundle $Q\to\bR$. 

\begin{prop} \label{gn1}  {\rm \cite{book,book99}.}
Each connection $\G$ on a 
bundle $Q\to\bR$ defines an atlas of local constant trivializations of
$Q\to\bR$ such that the associated bundle coordinates $(t,\rrq^i)$ on $Q$
possess the transition function
$\rrq^i\to \rrq'^i(\rrq^j)$ independent of $t$, and 
$\G=\dr_t$
with respect to these coordinates. Conversely, 
every atlas of local constant trivializations of the  bundle
$Q\to\bR$ sets a connection on  $Q\to\bR$ which is $\dr_t$
relative to this atlas.
\end{prop}

\begin{prop} {\rm \cite{book98,sard98}.}
Every bundle trivialization 
\beq
\psi: Q\cong  \bR\times M \label{m33}
\eeq 
 yields a complete
horizontal vector field $\G$ on this  bundle. Conversely, every complete
connection 
$\G$ on $Q\to\bR$ defines its trivialization (\ref{m33})
such that $\G=\dr_t$. 
\end{prop}

One can think 
of the atlas of
local constant trivializations and the bundle coordinates $(t,\rrq^i)$ in
Proposition \ref{gn1} as being also a reference frame corresponding to the 
connection $\G$.
These coordinates are said to be adapted to the reference frame $\G$. In
particular, the Hamiltonian form $H_\G$ relative to the adapted coordinates
reduces to the pure kinematic term $H_\G=p_idy^i$. Therefore, we will call
$H_\G$ a frame Hamiltonian form. 
Unless otherwise stated, by a reference frame will be meant
a complete reference frame. 
 Given a trivialization (\ref{m33}) of the
configuration bundle $Q\to\bR$, we have the corresponding trivializations of
velocity and momentum phase spaces
\be
J^1Q\cong \bR\times TM, \qquad V^*Q\cong \bR\times T^*M. 
\ee

\section{Interlude II. Lagrangian time-dependent dynamics}

 To obtain a complete
picture of the relations between Lagrangian and Hamiltonian time-dependent
mechanics in Section 6, we will refer to the following three types of PDEs in
the first order calculus of variations. These are Lagrange, Cartan and
Hamilton--De Donder equations.

Given a Lagrangian $L$ 
on the velocity phase space $J^1Q$, 
we follow the first variational formula of the
calculus of variations \cite{book,book98,sard97}, which provides the
canonical decomposition of the Lie derivative $\bL_{J^1u}L= (J^1u\rfloor\cL)dt$
of $L$ along a projectable vector field $u$ on $Q\to\bR$. We have 
\beq
J^1u\rfloor\cL= u_V\rfloor \cE_L + d_t(u\rfloor H_L), \label{C30} 
\eeq
where $u_V=(u\rfloor\th^i)\dr_i$, 
\beq
 H_L=L +\pi_i\th^i, \quad \pi_i=\dr^t_i\cL, 
\label{303}
\eeq
is the Poincar\'e--Cartan form and 
\beq
\cE_L=
 (\dr_i- d_t\pi_i)\cL \ol dq^i: J^2Q\to V^*Q \label{305}
\eeq
is the Euler--Lagrange operator associated with $L$. The kernel
$\Ker\cE_L\subset J^2Q$ of $\cE_L$ defines the Lagrange equations  on
$Q$, given by the coordinate relations
\beq
(\dr_i- d_t\pi_i)\cL=0. \label{b327} 
\eeq
On-shell,
the first variational formula (\ref{C30}) leads to 
the weak identity
\be
 \bL_{J^1u}L\ap d_t(u\rfloor H_L)dt, 
\ee
and then, if $\bL_{J^1u}L=0$,
to the weak conservation law 
\beq
0\ap dt(u\rfloor H_L)= - d_t \cT \label{K4}
\eeq
of the symmetry current
\beq
\cT =-(u\rfloor H_L)=-\pi_i(u^t q^i_t-u^i
)-u^t\cL. \label{Q30}
\eeq

Being the Lepagean equivalent of
the Lagrangian $L$ on $J^1Q$ (i.e., $L=h_0(H_L)$ where $h_0$ is the morphism
(\ref{cmp100})), the Poincar\'e--Cartan form $H_L$ (\ref{303}) is
also the Lepagean equivalent of the Lagrangian 
\beq
\ol L = \wh h_0(H_L) = (\cL + (\wh q_t^i - q_t^i)\pi_i)dt, \qquad \wh
h_0(dy^i)=\wh y^i_t dt, 
\label{cmp80}
\eeq
on the repeated jet manifold $J^1J^1Q$, coordinated by $(t,q^i,q^i_t,\wh
q^i_t, q^i_{tt})$. The 
Euler--Lagrange operator $\cE_{\ol L}:J^1J^1Q\to V^*J^1Q$ for $\ol L$ reads
\ben
&& \cE_{\ol L} = (\dr_i\cL - \wh d_t\pi_i 
+ \dr_i\pi_j(\wh q_t^j - q_t^j))\ol dq^i + \dr_i^t\pi_j(\wh
q_t^j - q_t^j) \ol dq_t^i, \label{2237} \\
&&\wh d_t=\dr_t +\wh q^i_t\dr_i +q^i_{tt}\dr_i^t. \nonumber
\een
Its kernel
$\Ker\cE_{\ol L}\subset J^1J^1Y$ defines the Cartan equations
\beq
\dr_i^t\pi_j(\wh q_t^j - q_t^j)=0, \qquad
\dr_i \cL - \wh d_t\pi_i 
+ (\wh q_t^j - q_t^j)\dr_i\pi_j=0. \label{b336}
\eeq
Since $\cE_{\ol L}\mid_{J^2Q}=\cE_L$, the Cartan equations (\ref{b336}) are
equivalent to the Lagrange equations  (\ref{b327}) on integrable
sections $\ol c=\dot c$ of
$J^1Q\to \bR$. These equations are equivalent 
in the case of regular Lagrangians.

On sections $\ol c: \bR\to J^1Q$, the Cartan equations (\ref{b336}) are
equivalent to the relation
\beq
 \ol c^*(u\rfloor dH_L)=0 \label{C28}
\eeq
which is assumed to hold for all vertical vector fields $u$ on $J^1Q\to \bR$.

With the Poincar\'e--Cartan form $H_L$ (\ref{303}), 
we have the Legendre morphism
\be
\wh H_L: J^1Q\op\to_Q T^*Q, \qquad 
(p_i, p)\circ\wh H_L =(\pi_i, \cL-\pi_i q^i_t ). 
\ee
Let $Z_L=\wh H_L(J^1Q)$  be an imbedded subbundle
$i_L:Z_L\hookrightarrow T^*Q$ of $T^*Q\to Q$.
It is provided with the pull-back 
 De Donder form $i^*_L\Xi$. We have 
\beq
H_L=\wh H_L^*\Xi_L=\wh H_L^*(i_L^*\Xi).  \label{cmp14}
\eeq
By analogy with the Cartan equations
(\ref{C28}), the  Hamilton--De Donder equations for sections $\ol r$ of 
$T^*Q\to \bR$ are written as
\beq
\ol r^*(u\rfloor d\Xi_L)=0 \label{N46}
\eeq
where $u$ is an arbitrary vertical vector field on
$T^*Q\to \bR$. 

\begin{theo}\label{ddd}  {\rm\cite{got91}.} Let the Legendre morphism
$\wh H_L:J^1Q\to Z_L$ be a submersion. Then a section $\ol c$ of $J^1Q\to \bR$
is a solution of the Cartan equations (\ref{C28}) iff $\wh
H_L\circ\ol c$ is a solution of the Hamilton--De Donder equations
(\ref{N46}), i.e., Cartan and Hamilton--De Donder equations are
quasi-equivalent.
\end{theo}

\section{Hamiltonian time-dependent dynamics}

Let the Legendre bundle $V^*Q\to\bR$ be provided with the holonomic
coordinates $(t,q^i,q^i_t)$. Relative to these coordinates, 
the canonical 3-form $\bom$ (\ref{z401}) and 
the canonical
Poisson structure (\ref{m72'}) on $V^*Q$  read
\ben
&& \bom = dp_i\w dq^i\w dt, \label{z401'} \\
&&\{f,g\}_V=\dr^if\dr_ig-\dr^ig\dr_if, \qquad f,g\in C^\infty{V^*Q}.
\label{m72}
\een 
The corresponding symplectic foliation  
coincides with the fibration $V^*Q\to \bR$. The symplectic forms on the
fibres of
$V^*Q\to\bR$ are the pull-backs $\Om_t=dp_i\w dq^i$
of the canonical symplectic form on the typical fibre $T^*M$ of the
Legendre bundle $V^*Q\to \bR$ with respect to trivialization morphisms
\cite{cari89,ham,sard98}. Given such a trivialization,
the Poisson structure (\ref{m72}) is isomorphic to the product of the
zero Poisson structure on $\bR$ and the canonical symplectic structure
on $T^*M$.

\begin{rem}
It is easily seen that an automorphism $\rho$ of the Legendre bundle
$V^*Q\to\bR$ is a canonical transformation of the Poisson structure
(\ref{m72}) iff it preserves the canonical 3-form $\bom$ (\ref{z401'}). 
Let us emphasize that canonical transformations are compatible with
the fibration $V^*Q\to\bR$, but not necessarily with the fibration 
$\pi_Q:V^*Q\to Q$.  We will restrict ourselves to the
holonomic coordinates on $V^*Y$ and holonomic transformations which are
obviously canonical.
\end{rem} 

With respect to the Poisson bracket (\ref{m72}), the Hamiltonian vector field  
$\vt_f$ for a function $f$ on the momentum phase space $V^*Q$ is 
\beq
\vt_f = \dr^if\dr_i- \dr_if\dr^i. \label{m73}
\eeq
A Hamiltonian vector field, by definition, is canonical.
A converse is the following.

\begin{prop} \label{ex0} 
Every vertical canonical
vector field on the Legendre bundle
$V^*Q\to\bR$ is locally a Hamiltonian vector field.
\end{prop}

The proof is based on the following facts. 

\begin{lem}\label{ex1} 
Let $\si$ be a 1-form on $V^*Q$. If $\si\w dt$ is closed form, it is exact.
\end{lem}

\begin{proof}
Since $V^*Q$ is diffeomorphic to $\bR\times T^*M$, we have the De
Rham cohomology group 
\be
H^2(V^*Q)= H^0(\bR)\ot H^2(T^*M) \oplus H^1(\bR)\ot H^1(T^*M).
\ee
The form $\si\w dt$ belongs to its second item which is zero.
\end{proof}

\begin{lem}\label{ex2} 
If the 2-form $\si\w dt$ is exact, then $\si\w dt= dg\w dt$ locally.
\end{lem}

\begin{proof}
The proof is based on the relative Poincar\'e lemma \cite{book}.
\end{proof} 

Let $\g=\dr_t +\g^i\dr_i +\g_i\dr^i$ be a canonical connection on the
Legendre bundle $V^*Q\to\bR$. Its components obey the relations  
\beq
\dr^i\g^j-\dr^j\g^i=0, \qquad
\dr_i\g_j- \dr_j\g_i=0,\qquad
\dr_j\g^i+\dr^i\g_j=0.\label{d025}
\eeq
Canonical connections constitute an affine space modelled
over the vector space of vertical canonical vector fields on $V^*Q\to \bR$.

\begin{prop}\label{ex3} 
If $\g$ is a canonical connection, then the form
$\g\rfloor\bom$ is exact. 
\end{prop}

\begin{proof}
Every connection $\G$ 
on $Q\to \bR$ gives rise to the connection
\beq
V^*\G=\dr_t +\G^i\dr_i -p_i\dr_j\G^i \dr^j \label{m38}
\eeq
on $V^*Q\to\bR$ 
which is a Hamiltonian connection for the frame Hamiltonian form
\beq
V^*\G\rfloor\bom=dH_\G, \qquad H_\G=p_idq^i -p_i\G^idt. \label{m61}
\eeq
Let us consider the decomposition $\g=V^*\G +\vt$,
where $\G$ is a connection on $Q\to \bR$. The assertion follows from  the
relation (\ref{m61}) and Proposition \ref{ex0}.
\end{proof}

Thus, every canonical connection $\g$ on $V^*Q$ defines an  
exterior 1-form $H$ modulo closed forms so that 
$dH=\g\rfloor\bom$. Such a form is called a locally Hamiltonian form.

\begin{prop}\label{germ} 
Every locally Hamiltonian form on the momentum phase space $V^*Q$ is locally a
Hamiltonian form modulo closed forms. 
\end{prop}

\begin{proof}
Given locally Hamiltonian forms $H_\g$ and $H_{\g'}$, their difference 
$\si=H_\g -H_{\g'}$ is a 1-form on $V^*Q$ such that the 2-form $\si\w dt$ is
closed. By virtue of Lemmas \ref{ex1} and \ref{ex2}, the form $\si\w dt$
is exact and $\si=fdt + dg$ locally. Put $H_{\g'}=H_\G$ where $\G$ is
a connection on $V^*Q\to\bR$. Then 
$H_\g$  modulo closed forms takes the local form $H_\g= H_\G + fdt$, and
coincides with the pull-back of the Liouville form
$\Xi$ on $T^*Q$ by the local section $p=-p_i\G^i+f$
of the fibre bundle (\ref{mm5}). 
\end{proof}

\begin{prop} \label{gm452'}  Conversely, each Hamiltonian form $H$
on the momentum phase space
$V^*Q$ admits a unique canonical connection $\g_H$ on $V^*Q\to \bR$
such that the relation (\ref{z405}) holds.
\end{prop}

\begin{proof} Given a Hamiltonian form $H$, its exterior differential 
\beq
dH=h^*d\Xi=(dp_i+\dr_i\cH dt)\w(dq^i-\dr^i\cH dt) \label{z408}
\eeq
is a presymplectic form of
constant rank $2m$ since the form 
\beq
(dH)^m =(dp_i\w dq^i)^m -m(dp_i\w dq^i)^{m-1}\w d\cH\w dt \label{m90}
\eeq
is nowhere vanishing. It is also seen that $(dH)^m\w dt\neq 0$.
It follows that the kernel of $dH$ is a 1-dimensional
distribution. Then the desired Hamiltonian connection 
\beq
\g_H=\dr_t +\dr^i\cH\dr_i -\dr_i\cH\dr^i \label{m57}
\eeq
is a unique vector field $\g_H$ on $V^*Q$  such that $\g_H\rfloor dH=0$,
$\g_H\rfloor dt=1$. 
\end{proof}

\begin{rem}
Hamiltonian forms constitute an affine space
modelled over the vector space of horizontal densities $fdt$ on $V^*Q\to\bR$,
i.e., over $C^\infty(V^*Q)$. Accordingly Hamiltonian
connections $\g_H$ form an affine space modelled over the vector 
space of Hamiltonian vector fields.
Every Hamiltonian form $H$ defines the associated Hamiltonian map 
\beq
\wh H=J^1\pi_Q\circ\g_H:\dr_t + \dr^i\cH:V^*Q\to J^1Q. \label{mm41}
\eeq
With the Hamiltonian
map (\ref{mm41}), we have another Hamiltonian form
\beq
H_{\wh H}= -\wh H\rfloor\bth= p_idq^i -p_i\dr^i\cH. \label{mm42}
\eeq
It is readily observed that $H_{\wh H}=H$ iff $H$ is a frame Hamiltonian
form.
\end{rem}

Given a Hamiltonian connection $\g_H$ (\ref{m57}), the corresponding Hamilton
equations 
$D_{\g_H}=0$ take the coordinate form
\bea
&& q^i_t =\dr^i\cH, \label{m41a}\\
&&  p_{ti} =-\dr_i\cH. \label{m41b}
\eea
Their classical solutions are integral sections of the
Hamiltonian connection $\g_H$, i.e.,
$\dot r=\g_H\circ r$.
On sections 
$r$ of the Legendre bundle $V^*Q\to \bR$, the Hamilton equations (\ref{m41a})
-- (\ref{m41b}) 
are equivalent to the relation
\beq
r^*(u\rfloor dH)=0 \label{y5}
\eeq
which is assumed to hold for any vertical vector field $u$ on $V^*Q\to\bR$.

The following two constructions are useful.

It is readily observed that a Hamiltonian form $H$
(\ref{b4210}) is the Poincar\'e--Cartan form (\ref{303}) for the Lagrangian 
\beq
L_H=h_0(H) = (p_iq^i_t - \cH)\om \label{Q3}
\eeq
on the jet manifold $J^1V^*Q$.  Given
a projectable vector field $u$ on the configuration bundle
$Q\to\bR$ and its lift
\beq
\wt u=u^t\dr_t + u^i\dr_i - \dr_i u^j p_j\dr^i \label{cmp90}
\eeq
onto the Legendre bundle $V^*Q\to\bR$, we have
\beq
\bL_{\wt u}H= \bL_{J^1\wt u}L_H. \label{b4180}
\eeq
It is easily seen that the
Hamilton equations (\ref{m41a}) -- (\ref{m41b}) for $H$
are exactly the Lagrange equations
for $L_H$, i.e., they 
characterize  the
kernel of the Euler--Lagrange operator 
\beq
 \cE_H= (q^i_t-\dr^i\cH)\ol dp_i -(p_{ti}+\dr_i\cH)\ol dq^i: J^1V^*Q\to
V^*V^*Q \label{mm40}
\eeq
for the Lagrangian $L_H$, called the Hamilton
operator for $H$.  

Using the relation (\ref{b4180}), let us obtain the Hamiltonian conservation
laws in time-dependent mechanics. As in field theory, by gauge
transformations in time-dependent mechanics are meant automorphism of the
configuration bundle $Q\to \bR$, but only over translations of
the base $\bR$. Then, projectable vector fields 
\beq
u=u^t\dr_t +u^i\dr_i, \qquad u\rfloor dt=u^t={\rm const.}, \label{m223}
\eeq
on $V^*Q\to\bR$ can be seen as generators of local 1-parameter groups
of local gauge transformations.
Given a Hamiltonian form $H$ (\ref{b4210}), its Lie derivative (\ref{b4180})
reads
\beq
\bL_{\wt u}H= \bL_{J^1\wt u}L_H=(-u^t\dr_t\cH+p_i\dr_tu^i -u^i\dr_i\cH +\dr_j
u^ip_i\dr^j\cH)dt. \label{mm24}
\eeq
The first variational formula (\ref{C30}) applied to the Lagrangian
$L_H$ (\ref{Q3}) leads to the weak identity $\bL_{\wt u}H\ap d_t(u\rfloor
H)dt$. If the Lie derivative (\ref{mm24}) vanishes, we have the conserved
symmetry current
\beq
J_u= u\rfloor dH= p_iu^i - u^t\cH, \label{b4306}
\eeq
along $u$. Every vector field (\ref{m223}) is a
superposition of a vertical vector field and a reference frame on $Q\to\bR$.
If $u$ is a vertical vector field, $J_u$ is the N\"other current 
\beq
J_u(q)= u\rfloor q=p_iu^i, \qquad q=p_i\ol dq^i\in V^*Q. \label{z606}
\eeq
The symmetry current along a reference frame $\G$
\beq
J_\G=p_i\G^i-\cH=-\wt\cH_\G \label{mm25}
\eeq
is the energy function
with respect to the reference frame $\G$, taken with the sign minus
\cite{eche95,book98,sard98}. It is readily observed that, given a Hamiltonian
form $H$, the energy functions $\wt\cH_\G$ constitute an affine space modelled
over the vector space of N\"other currents. 

\begin{prop} \label{mm30}
Given a Hamiltonian form $H$, the conserved currents (\ref{b4306})
form a Lie algebra with respect to the Poisson bracket
\beq
\{J_u,J_{u'}\}_V=J_{[u,u']}. \label{c2}
\eeq
\end{prop}

The second of the above mentioned constructions enables us to represent
the right-hand side of the evolution equation (\ref{mm15}) as a pure Poisson
bracket. Given a Hamiltonian form $H=h^*\Xi$, let us consider its pull-back
$\zeta^*H$ onto the cotangent bundle $T^*Q$. It is readily observed that the
difference
$\Xi-\z^*H$ is a horizontal 1-form on $T^*Q\to\bR$, while  
\beq
\cH^*=\dr_t\rfloor(\Xi-\zeta^*H)=p+\cH \label{mm16}
\eeq
is a function on $T^*Q$. Then the relation
\beq
\z^*(\bL_{\g_H}f)=\{\cH^*,\z^*f\}_T \label{mm17}
\eeq
holds for every function $f\in C^\infty(V^*Q)$. 
In particular, given a projectable vector
field $u$ (\ref{m223}), the symmetry current $J_u$ (\ref{b4306}) is conserved
if and only if 
\beq
\{\cH^*,\z^*J_u\}_T=0. \label{mm31}
\eeq
Moreover, let $\vt_{\cH^*}$
be the Hamiltonian vector field for the function $\cH^*$ (\ref{mm16}) with
respect to the canonical Poisson structure $\{,\}_T$ on $T^*Q$. Then
\beq
T\z(\vt_{\cH^*})=\g_H. \label{mm18}
\eeq

\section{Time-dependent constraints}

As was mentioned above, an algebra of time-dependent constraints on
the momentum phase space $V^*Q$ can be described similarly to that in
conservative Hamiltonian mechanics.

Let $N$ be a closed imbedded subbundle $i_N:N\op\hookrightarrow V^*Q$ of the
Legendre bundle
$V^*Q\to\bR$, treated as a constraint space.  Let us consider the ideal
$I_N$ of real functions
$f$ on
$V^*Q$ which vanish on $N$, i.e.,
$i_N^*f=0$. Its elements are said to be constraints. There is the isomorphism 
\beq
C^\infty(V^*Q)/I_N\cong C^\infty(N)\label{gm93}
\eeq 
of associative commutative algebras. $N$ cannot be
neither Lagrangian nor symplectic submanifold with respect to the Poisson
structure on $V^*Q$.
By the normalize $\ol I_N$ of the ideal
$I_N$ is meant the subset of functions of $C^\infty(V^*Q)$ whose Hamiltonian
vector fields restrict to vector fields on $N$, i.e., 
\beq
\ol I_N=\{f\in C^\infty(V^*Q):\,\, \{f,g\}_V\in I_N, \,\, \forall g\in I_N\},
\label{gm95}
\eeq
\cite{kimura}. It follows from the Jacobi identity that the normalizer
(\ref{gm95}) is a Poisson subalgebra of
$C^\infty(V^*Q)$. Put
\beq
I'_N= \ol I_N\cap I_N. \label{gm96}
\eeq
It is naturally a Poisson subalgebra of $\ol I_N$. Its elements are called 
the first class constraints, while
the remaining elements of $I_N$ are the second class constraints. It is
readily observed that $I^2_N\subset I'_N$, i.e., the products of
second class constraints are first class constraints.

\begin{rem}
Let $N$ be a coisotropic
submanifold of $V^*Q$, i.e., $w^\sh(\rA TN)\subset TN$. Then $I_N\subset
\ol I_N$ and $I_N=I'_N$, i.e., all constraints are of the first class.
\end{rem}

The relation (\ref{mm17}) enables us to extend the constraint algorithm of
conservative mechanics and time-dependent mechanics on a product $\bR\times
M$ (see \cite{chinea,leon93}) to mechanical systems subject to time-dependent
transformations. 

Let $H$ be a Hamiltonian form on the momentum phase space $V^*Q$.
In accordance with the relation (\ref{mm17}), a constraint $f\in I_N$ is
preserved if the bracket (\ref{mm17}) vanishes. It follows that the
solutions of the Hamilton equations (\ref{m41a}) -- (\ref{m41b}) do not leave
the constraint space $N$ if 
\beq
\{\cH^*,\z^*I_N\}_T\subset \z^*I_N. \label{mm20}
\eeq
If the relation (\ref{mm20}) fails to hold, let us introduce  
secondary constraints $\{\cH^*,\zeta^*f\}_T$, $f\in I_N$, which belong to
$\zeta^* C^\infty(V^*Q)$. If the collection of primary and secondary
constraints is not closed  with respect to the relation (\ref{mm20}), let us
add the tertiary constraints $\{\cH^*,\{\cH^*,\zeta^*f_a\}_T\}_T$ and so
on.

Let us assume that $N$ is a final constraint space for a Hamiltonian form
$H$.  If a Hamiltonian form $H$ satisfies the relation (\ref{mm20}), so is a
Hamiltonian form
\beq
H_f=H-fdt \label{mm21}
\eeq
where $f\in I'_N$ is a first class constraint. 
Though Hamiltonian forms $H$ and $H_f$ coincide
with each other on the constraint space $N$, the corresponding Hamilton
equations have different solutions on the constraint space $N$ because
$dH\mid_N\neq dH_f\mid_N$. At the same time, $d(i_N^*H)=d(i_N^*H_f)$.
Therefore, let us introduce the constrained Hamiltonian form
\beq
H_N=i_N^*H_f \label{mm23}
\eeq
which is the same for all $f\in I'_N$.
Note that $H_N$ (\ref{mm23}) is not a true Hamiltonian form on
$N\to \bR$ in general.
On sections $r$ of the fibre bundle $N\to\bR$, we can write the equations
\beq
r^*(u_N\rfloor dH_N) =0, \label{N44}
\eeq
where $u_N$ is an arbitrary vertical vector field on $N\to \bR$. They are
called the constrained
Hamilton equations. 

\begin{prop} \label{cmp22}  For any Hamiltonian form $H_f$
(\ref{mm21}), every solution
of the Hamilton equations which lives in the constraint space
$N$ is a solution of the constrained Hamilton equations (\ref{N44}).
\end{prop}

\begin{proof} The constrained Hamilton
equations can be written as
\beq
r^*(u_N\rfloor di^*_NH_f)=r^*(u_N\rfloor dH_f\mid_N) =0. \label{N44'}
\eeq
They differ from the Hamilton equations (\ref{y5}) for $H_f$ restricted to $N$
which read
\beq
r^*(u\rfloor dH_f\mid_N) =0, \label{cmp10}
\eeq
where $r$ is a section of $N\to \bR$ and $u$ is an arbitrary vertical vector
field on $V^*Q\to \bR$. A solution $r$ of the equations
(\ref{cmp10}) satisfies obviously the weaker condition
(\ref{N44'}).
\end{proof}

\begin{rem}
One also can consider the problem  of constructing a generalized Hamiltonian
system, similar to that for Dirac constraint system in conservative mechanics
\cite{book98}. Let $H$ satisfies the condition
$\{\cH^*,\zeta^*I'_N\}_T\subset I_N$, whereas
$\{\cH^*,\zeta^*I'_N\}_T\not\subset I_N$. The goal is to find 
a constraint $f\in I_N$  such that the modified Hamiltonian $H -fdt$ would
satisfy both the conditions
\be
\{\cH^* +\zeta^*f ,\zeta^* I'_N\}_T\subset \zeta^*I_N, \qquad 
\{\cH^* +\zeta^*f ,\zeta^* I_N\}_T\subset \zeta^*I_N.
\ee
The first of them is fulfilled for any $f\in I_N$,
while the latter is an equation for a second-class constraint
$f$. 
\end{rem}

It should be emphasized that, in contrast with the conservative
case, the Hamiltonian vector fields $\vt_f$ for the first class constraints
$f\in I'_N$ in time-dependent mechanics are not generators of gauge symmetries
of a Hamiltonian form in general. At the same time, generators of gauge
symmetries define an ideal of constraints as follows.

The above construction, except the isomorphism
(\ref{gm93}), can be applied to any ideal $I$ of $C^\infty(V^*Q)$. Then one
says that the Poisson algebra $\ol I/ I'$ is the reduction of the Poisson
algebra
$C^\infty(V^*Q)$ via the ideal $I$ \cite{kimura}.
In particular, an ideal $I$ is said to be
coisotropic if it is a Poisson algebra. In this case, $I$ is a Poisson
subalgebra of the normalize
$\ol I$ (\ref{gm95}), and coincides with $I'$ (\ref{gm96}).

Let $\cA$ be a Lie algebra of generators $u$ of gauge symmetries of a
Hamiltonian form $H$. In accordance with the relation (\ref{c2}), the
corresponding symmetry currents $J_u$ (\ref{b4306}) on
$V^*Q$ constitute a Lie algebra with
respect to the canonical Poisson bracket on $V^*Q$. Let $I_\cA$ denotes the
ideal of $C^\infty(V^*Q)$ generated by these symmetry currents. 
It is readily observed that this ideal is coisotropic. Then one can think of
$I_\cA$ as being an ideal of first class constraints compatible with the
Hamiltonian form $H$, i.e., 
\beq
\{\cH^*,\zeta^*I_\cA\}_T\subset \zeta^*I_\cA. \label{mm33}
\eeq
Note that any Hamiltonian form $H_u=H-J_udt$,
$u\in\cA$, obeys the same relation (\ref{mm33}), but other currents $J_{u'}$
are not conserved with respect $H_u$ if $[u,u']\neq 0$.

 Let now $\cA$ be an arbitrary Lie algebra of vertical vector fields $u$ on the
configuration bundle $Q\to\bR$. The relation (\ref{c2}) remains true, while
the corresponding symmetry currents $J_u$ (\ref{z606}) on
$V^*Q$ constitute a Lie algebra and generate the corresponding coisotropic
ideal $I_\cA$ of $C^\infty(V^*Q)$ with respect to the canonical Poisson bracket
on $V^*Q$.

\begin{prop}
Let $\cA$ be a finite-dimensional Lie algebra of vertical vector fields on
the configuration bundle $Q\to\bR$. If there exists a reference frame $\G$ on
$Q\to\bR$ such that $[\G,\cA]=0$, then there exists a non-frame Hamiltonian
form $H$ on the Legendre bundle $V^*Q$ such that $\cA$ is the algebra of gauge
symmetries of $H$. 
\end{prop} 

\begin{proof}
Let $\ol\cA$ be the universal enveloping algebra of the Lie algebra of the
symmetry currents $J_u$, $u\in \cA$, (\ref{z606}). Then each non-zero
element $C$ of its center of order $>1$ can be written as a polynomial in
$J_u$, and defines the desired Hamiltonian form $H=H_\G -Cdt$.
\end{proof}

\section{Lagrangian constraints}

Let us consider the Hamiltonian description of Lagrangian mechanical systems
on a configuration bundle $Q\to\bR$. If a Lagrangian is degenerate, we have
the Lagrangian constraint subspace of the Legendre bundle $V^*Q$ and a set of
Hamiltonian forms associated with the same Lagrangian. 
Given a Lagrangian $L$ (\ref{mm43}) on the velocity phase space $J^1Q$,
a Hamiltonian form $H$ on the momentum phase space $V^*Q$ is said to be
associated with $L$ if $H$ satisfies the relations
\bea
&&\wh L\circ\wh H\circ \wh L=\wh L,\label{2.30a} \\
&&H=H_{\wh H}+\wh H^*L \label{2.30b}
\eea
where $\wh H$ and $\wh L$ are the Hamiltonian morphism (\ref{mm41})
and the Legendre map (\ref{a303}), respectively.  A glance
at the relation (\ref{2.30a}) shows that $\wh L\circ\wh H$ is the projector
\beq
p_i(z)=\pi_i(t,q^i,\dr^j\cH(z)), \qquad z\in N_L,
\label{b481'}
\eeq
from $\Pi$ onto the Lagrangian constraint space $N_L=\wh L( J^1Y)$.
Accordingly,  $\wh H\circ\wh L$ is the projector from $J^1Y$ onto $\wh
H(N_L)$.  A Hamiltonian form is called weakly associated with a Lagrangian $L$
if the condition (\ref{2.30b}) holds on the Lagrangian constraint space $N_L$.

\begin{prop} \label{jp}  {\rm \cite{book}.}
If a bundle morphism $\Phi:V^*Q\op\to_Q J^1Q$ obeys the relation
(\ref{2.30a}), then the Hamiltonian form $H=-\Phi\rfloor\bth+\Phi^*L$ is
weakly associated with the Lagrangian $L$. If $\Phi=\wh H$, then $H$ is
associated with $L$.
\end{prop}

\begin{lem} \label{cmp110} 
Any Hamiltonian form $H$ weakly associated with a Lagrangian $L$ obeys
the relation
\beq
H\mid_{N_L}=\wh H^*H_L\mid_{N_L}, \label{4.9}
\eeq
where $H_L$ is the Poincar\'e--Cartan form (\ref{303}).
\end{lem}

\begin{proof}
The relation (\ref{2.30b}) takes the coordinate form
\beq
\cH(z)=p_i\dr^i\cH-\cL(t,q^i,\dr^j\cH(z)), \qquad z\in N_L. \label{b481}
\eeq
Substituting (\ref{b481'}) and (\ref{b481}) in (\ref{b4210}), we
obtain the relation (\ref{4.9}).
\end{proof}

The difference between associated and weakly associated Hamiltonian forms
lies in the following. Let $H$ be an associated Hamiltonian form, i.e., the
equality (\ref{b481}) holds everywhere on $V^*Q$.
The exterior differential of this
equality leads to the relations
\be
&& \dr_t\cH(z) =-(\dr_t\cL)\circ \wh H(z), \qquad
\dr_i\cH(z) =-(\dr_i\cL)\circ \wh H(z), \qquad
z\in N_L, \\ 
&&
(p_i-(\dr^t_i\cL)(t,q^i,\dr^j_t\cH))\dr^i_t\dr^a_t\cH=0.
\ee
The last of them shows that the Hamiltonian form is not regular outside
the Lagrangian constraint space $N_L$. In particular, any Hamiltonian form is
weakly associated with the Lagrangian $L=0$, while the associated Hamiltonian
forms are only $H_\G$.

Here we restrict our consideration to almost regular Lagrangians $L$, i.e.,
if: (i) the Lagrangian constraint space $N_L$ is a closed imbedded subbundle
$i_N:N_L\to V^*Q$ of the bundle $V^*Q\to Q$, (ii) the Legendre map $\wh
L:J^1Q\to N_L$ is a fibred manifold, and (iii) the pre-image $\wh L^{-1}(z)$
of  any point $z\in N_L$ is a connected submanifold of $J^1Q$.

\begin{prop} \label{mm71} 
As an immediate consequence of the above conditions (i), (ii) and
Proposition
\ref{jp}, a Hamiltonian form $H$ weakly associated with an almost regular
Lagrangian $L$ exists iff the fibred manifold $J^1V^*Q\to N_L$ admits a global
section.
\end{prop}

The condition (iii) leads to the following property.

\begin{lem} \label{3.22}  {\rm \cite{book,book98}.}
The 
Poincar\'e--Cartan form $H_L$ for an almost regular 
 Lagrangian $L$ is constant on the connected
pre-image $\wh L^{-1}(z)$ of any point  $z\in N_L$.
\end{lem}

An immediate consequence of this fact is the following assertion.

\begin{prop} \label{3.22'}  {\rm \cite{book}.} All Hamiltonian forms
weakly associated with an almost regular Lagrangian $L$ coincide with each
other on the Lagrangian constraint space $N_L$,
and the Poincar\'e--Cartan form $H_L$ (\ref{303})
for $L$ is the pull-back
\beq
H_L=\wh L^*H, \qquad
 \pi_iq^i_t-\cL=\cH(t,q^j,\pi_j), \label{2.32}
\eeq
of any such a Hamiltonian form $H$.
\end{prop}

It follows that, given Hamiltonian forms $H$ an $H'$ weakly associated with an
almost regular Lagrangian $L$, their difference is $fdt$, $f\in I_N$.
However, $\wh H\mid_{N_L}\neq \wh H'\mid_{N_L}$ in general. Therefore, the
Hamilton equations for $H$ and $H'$ do not coincide necessarily on the
Lagrangian constraint space $N_L$. Their solutions can
leave $N_L$, i.e., the relation (\ref{mm20}) fails to hold in general.

Proposition \ref{3.22'} enables us to connect Lagrange and Cartan
equations for an almost regular Lagrangian $L$ with the Hamilton
equations for Hamiltonian forms weakly associated with $L$ \cite{book}.

\begin{theo}\label{3.23} 
Let a section $r$ of $V^*Q\to \bR$
be a  solution of the Hamilton equations (\ref{m41a}) -- (\ref{m41b})
for a Hamiltonian form $H$ weakly associated with an almost regular
Lagrangian $L$. If $r$ lives in the constraint space $N_L$, the
section $c=\pi_Q\circ r$
of $Q\to \bR$ satisfies the Lagrange
equations (\ref{b327}), while $\ol c=\wh H\circ r$ obeys the Cartan equations
(\ref{b336}). 
\end{theo}

The proof is based on the relation 
\beq
\cE_{\ol L}=(J^1\wh L)^*\cE_H \label{b4.1000}
\eeq
or on the
equivalent relation $\ol L=(J^1\wh L)^*L_H$ which are derived from the
equality (\ref{2.32}). The converse assertion is more intricate.

\begin{theo}\label{3.24}  Given an almost regular Lagrangian $L$,
let a section $\ol c$ of the jet bundle
$J^1Q\to \bR$ be a solution of the
Cartan equations (\ref{b336}).
Let $H$ be a Hamiltonian form weakly associated with $L$,  and let $H$ satisfy
the relation
\beq
\wh H\circ \wh L\circ \ol c=J^1(\pi^1_0\circ\ol c).\label{2.36}
\eeq
Then, the section $r=\wh L\circ \ol c$
of the Legendre bundle $V^*Q\to \bR$ is a solution of the
Hamilton equations (\ref{m41a}) -- (\ref{m41b}) for $H$.
\end{theo}

\begin{rem} \label{cmp9} 
Since $\wh H\circ \wh L$ in Theorem (\ref{3.24}) is a projection operator,
the condition (\ref{2.36}) implies that the solution $\ol s$ of the Cartan
equations is actually an integrable section $\ol c=\dot c$ where $c$ is a
solution of the Lagrange equations.  
In fact, the
relation (\ref{b4.1000}) gives more than it is needed for proving Theorem
\ref{3.23}. Using this relation, one can justify that, if $\g$ is a Hamiltonian
connection for a Hamiltonian form $H$ weakly associated with an almost regular
Lagrangian
$L$, then the composition $J^1\wh H\circ\g\circ\wh L$ takes its values in
$\Ker \cE_{\ol L}\cap J^2Y$, i.e., this is a local holonomic Lagrangian
connection on
$\wh H(N_L)$ \cite{book}.  A converse of this assertion, however, fails to
be true in the case of degenerate Lagrangians. 
Let a Lagrangian $L$ be
hyperregular, i.e., the Legendre map $\wh L$
is a diffeomorphism. Then $\wh L^{-1}$ is a Hamiltonian map, and there is a
unique Hamiltonian form $H=H_{\wh L^{-1}}+\wh L^{-1*}L$
weakly associated with $L$. In this case, both the relation (\ref{b4.1000}) and
the converse one $\cE_H=(J^1\wh H)^*\cE_{\ol L}$
hold. It follows that the Lagrange equations for $L$ and the Hamilton
equations for $H$ are equivalent.
\end{rem}

We will say that a set of Hamiltonian forms
$H$ weakly associated with an almost regular Lagrangian $L$ is
complete if, for each
solution
$c$ of the Lagrange equations, there exists a solution
$r$ of the Hamilton equations for a Hamiltonian form $H$ from this 
set such that $c=\pi_Q\circ r$.
By virtue of Theorem \ref{3.24} and Remark \ref{cmp9}, a set
of weakly associated Hamiltonian forms
is complete if, for every solution $c$ on $\bR$ of the Lagrange
equations for $L$, there is a Hamiltonian form $H$ from this
set which fulfills the relation 
\beq
\wh H\circ \wh L\circ \dot c=\dot c. \label{2.36'}
\eeq

In accordance with Proposition \ref{mm71}, on an open neighbourhood in $V^*Q$
of each point $z\in N_L$, there exists a complete set of local Hamiltonian
forms weakly associated with an almost regular Lagrangian 
$L$. Moreover, one can always construct a complete set of associated local
Hamiltonian forms \cite{sard95,zak}

Given a Hamiltonian form $H$ weakly associated with an almost regular
Lagrangian $L$, let us consider the corresponding constrained Hamiltonian
form $H_N$ (\ref{mm23}). By virtue of
Proposition  (\ref{3.22'}),
$H_N$ is the same for all Hamiltonian forms weakly associated with $L$, and
$H_L=\wh L^* H_N$.  The first of these facts leads to the assertion proved
similarly to Proposition \ref{cmp22}.

\begin{prop} \label{mm72}   For any Hamiltonian form $H$ weakly
associated with an almost regular Lagrangian $L$, every solution
 of the Hamilton equations which lives in the Lagrangian constraint space
$N_L$ is a solution of the constrained Hamilton equations (\ref{N44}).
\end{prop}

Using the equality $H_L=\wh L^* H_N$, one can show that the constrained
Hamilton equations (\ref{N44}) are equivalent to the
Hamilton--De Donder equations (\ref{N46}) and, by virtue of Theorem
\ref{ddd}, are quasi-equivalent to the Cartan equations
(\ref{C28}) \cite{book,book98}.

\section{Quadratic degenerate systems}

Let us study the important case of almost regular
quadratic Lagrangians. We show that, in this case, there always exist both a
complete set of associated Hamiltonian forms and a complete set of
non-degenerate weakly associated Hamiltonian forms. The latter is important
for quantization.

Given a configuration bundle
$Q\to \bR$, let us consider a  quadratic Lagrangian $L$ which has the
coordinate expression
\beq
\cL=\frac12 a_{ij} q^i_t q^j_t +
b_i q^i_t + c, \label{N12}
\eeq
where $a$, $b$ and $c$ are local functions on $Q$. This property is
coordinate-independent due to the affine transformation law of the coordinates
$q^i_t$. The associated Legendre map 
\beq
p_i\circ\wh L= a_{ij} q^j_t +b_i \label{N13}
\eeq
is an affine morphism over $Q$. It defines the corresponding linear
morphism
\beq
\ol L: VQ\op\to_Q V^*Q,\qquad p_i\circ\ol
L=a_{ij}\dot q^j. \label{N13'}
\eeq

Let the Lagrangian $L$ (\ref{N12}) be almost regular, i.e.,
the matrix function $a_{ij}$ is of constant rank. Then
the Lagrangian constraint space $N_L$ 
(\ref{N13}) is an affine subbundle of the bundle $V^*Q\to Q$, modelled
over the vector subbundle $\ol N_L$ (\ref{N13'}) of  $V^*Q\to Q$. 
Hence, $N_L\to Q$ has a global section. For the sake of simplicity, let us
assume that it is the canonical
zero section $\wh 0(Q)$ of $V^*Q\to Q$. Then $\ol N_L=N_L$.
Accordingly, the kernel
of the Legendre map (\ref{N13})  is an affine
subbundle of the affine jet bundle $J^1Q\to Q$, modelled over the kernel of
the linear morphism $\ol L$ (\ref{N13'}). Then there exists a connection 
\ben
&&\G: Q\to \Ker\wh L\subset J^1Q, \label{N16}\\
&& a_{ij}\G^j_\m + b_i =0, \label{250}
\een
on $Q\to \bR$.
Connections (\ref{N16}) constitute an affine space modelled over the linear
space of vertical vector fields $\up$ on $Q\to \bR$, satisfying the conditions
\beq
a_{ij}\up^j =0 \label{cmp21}
\eeq
and, as a consequence, the conditions $\up^i b_i=0$.
If the Lagrangian (\ref{N12}) is regular, the
connection (\ref{N16}) is unique.

The matrix $a$ in the Lagrangian $L$ (\ref{N12}) can be seen as a degenerate
fibre metric of constant rank in 
$VQ\to Q$. Then it satisfies the following Lemma.

\begin{lem} \label{mm45} 
Given a $k$-dimensional vector bundle $E\to Z$, let $a$ be a section of rank
$r$ of the tensor bundle $\op\vee^2E^*\to Z$. There is a splitting 
\beq
E= \Ker a\op\oplus_Z  E' \label{mm50}
\eeq
where $E'=E/\Ker a$ is the quotient bundle, and $a$ is a non-degenerate
fibre metric in $E'$.
\end{lem} 

\begin{proof}
Since $a$ exists, the structure group $GL(k,\bR)$ of the vector bundle $E\to
Z$ is reducible to the subgroup $GL(r,k-r;\bR)$ of general linear
transformations of $\bR^k$ which keep its $r$-dimensional subspace, and to
its subgroup $GL(r,\bR)\times GL(k-r,\bR)$.
\end{proof}

\begin{theo}\label{04.2}  There exists a linear bundle
map
\beq
\si: V^*Q\op\to_Q VQ, \qquad
\dot q^i\circ\si =\si^{ij}p_j, \label{N17}
\eeq
such that $\ol L\circ\si\circ i_N= i_N$.
\end{theo}

\begin{proof} 
The map (\ref{N17}) is a solution of the algebraic equations
\beq
a_{ij}\si^{jk}a_{kb}=a_{ib}. \label{mm100}
\eeq
By virtue of Lemma \ref{mm45}, there exist the bundle slitting 
\beq
VQ=\Ker a\op\oplus_Q E' \label{mm46}
\eeq
and a (non-holonomic) atlas of this bundle such that transition
functions  of
$\Ker a$ and $E'$ are independent. Since $a$ is a non-degenerate fibre metric
in $E'$, there exists an atlas of $E'$ such
that $a$ is brought into a diagonal matrix with non-vanishing
components
$a_{AA}$. Due to the splitting (\ref{mm46}), we have the corresponding
bundle splitting
\beq
V^*Q=(\Ker a)^*\op\oplus_Q \im a. \label{mm46'}
\eeq
Then the desired map $\si$ is represented by a direct sum $\si_1\oplus\si_0$
of an arbitrary section $\si_1$ of the bundle
$\op\vee^2\Ker a^*\to Q$
and the section
$\si_0$ of the bundle $\op\vee^2E'\to Q$,
which has non-vanishing components $\si^{AA}=(a_{AA})^{-1}$ with respect to
the above mentioned atlas of $E'$.  Moreover,
$\si$ satisfies the particular relations
\beq
\si_0=\si_0\circ\ol L\circ\si_0, \quad a\circ\si_1=0, \quad \si_1\circ a=0.
\label{N21}
\eeq
\end{proof}

\begin{cor} 
The splitting (\ref{mm46}) leads to the splitting 
\bea
&& J^1Q=\cS(J^1Q)\op\oplus_Q \cF(J^1Q)=\Ker\wh L\op\oplus_Q{\rm Im}(\si\circ
\wh L), \label{N18} \\
&& q^i_t=\cS^i+\cF^i= [q^i_t
-\si^{ik}_0 (a_{kj}q^j_t + b_k)]+
[\si^{ik}_0 (a_{kj}q^j_t + b_k)], \label{b4122}
\eea
while the splitting (\ref{mm46'})
can be written as
\bea
&& V^*Q=\cR(V^*Q)\op\oplus_Q\cP(V^*Q)=\Ker\si_0 \op\oplus_Q N_L, \label{N20} \\
&& p_i = \cR_i+\cP_i= [p_i -
a_{ij}\si^{jk}_0p_k] +
[a_{ij}\si^{jk}_0p_k]. \label{N20'}
\eea
\end{cor}

It is readily observed that, with respect to the coordinates $\cS^i_\la$
and $\cF^i_\la$ (\ref{b4122}), the Lagrangian (\ref{N12}) reads 
\beq
\cL=\frac12 a_{ij}\cF^i\cF^j +c', \label{cmp31}
\eeq
while the Lagrangian constraint space is given by the reducible constraints
\beq
\cR_i= p_i -
a_{ij}\si^{jk}_0p_k=0. \label{zzz}
\eeq

Given the linear map $\si$ (\ref{N17}) and the connection $\G$
(\ref{N16}), let us consider the affine Hamiltonian map
\beq
\Phi=\wh\G+\si:V^*Q \op\to J^1Q,  \qquad
\Phi^i = \G^i  + \si^{ij}p_j, \label{N19}
\eeq
and the Hamiltonian form
\ben
&& H=H_\Phi +\Phi^*L= p_idq^i - [p_i\G^i
 +\frac12 \si_0{}^{ij}p_ip_j
+\si_1{}^{ij}p_ip_j -c']dt=
\label{N22}\\
&& \qquad (\cR_i+\cP_i)dq^i - [(\cR_i+\cP_i)\G^i
+\frac12
\si_0^{ij}\cP_i\cP_j
+\si_1^{ij}p_ip_j -c']dt.\nonumber
\een
In particular, if $\si_1$ is non-degenerate, so is the Hamiltonian form
(\ref{N22}). 

\begin{theo} \label{cmp30}  The Hamiltonian forms (\ref{N22})
parameterized by connections $\G$ (\ref{N16}) are weakly associated with the
Lagrangian  (\ref{N12}) and constitute a complete set.
\end{theo}

\begin{proof}
By the very definitions of $\G$ and $\si$, the Hamiltonian map (\ref{N19})
satisfies the condition (\ref{2.30a}). Then $H$ is weakly associated with $L$
(\ref{N12}) in accordance with Proposition \ref{jp}. 
Let us write the corresponding Hamilton equations (\ref{m41a}) for
a section $r$ of the Legendre bundle $V^*Q\to \bR$. They are
\beq
\dot c= (\wh\G+\si)\circ r, \qquad c=\pi_Q\circ r. \label{N29}
\eeq
Due to the surjections $\cS$ and $\cF$ (\ref{N18}),
the Hamilton equations (\ref{N29}) break in two parts
\ben
&&\cS\circ \dot c=\G\circ c, \qquad \dot r^i-
\si^{ik} (a_{kj}\dot r^j + b_k)=\G^i\circ c,
\label{N23} \\
&&\cF \circ \dot c=\si\circ r, \qquad
\si^{ik} (a_{kj}\dot r^j + b_k)=
\si^{ik}r_k.\label{N28}
\een
Let $c$ be an arbitrary section of $Q\to \bR$,
e.g., a solution of the Lagrange
equations. There exists a connection $\G$ (\ref{N16}) such
that the relation (\ref{N23}) holds, namely, $\G={\cal S}\circ\G'$ where
$\G'$ is a
connection on $Q\to \bR$ which has $c$ as an integral section. 
It is easily seen that, in this case, the Hamiltonian map (\ref{N19})
satisfies the relation (\ref{2.36'}) for $c$. 
Hence, the Hamiltonian forms (\ref{N22}) constitute
a complete set. 
\end{proof}

It is readily observed that, if $\si_1=0$, then $\Phi=\wh H$ and the
Hamiltonian forms (\ref{N22}) are associated with the Lagrangian (\ref{N12})
in accordance with Proposition \ref{jp}. Thus, for different $\si_1$, we have
different complete sets of Hamiltonian forms (\ref{N22}). Hamiltonian forms $H$
(\ref{N22}) of such a complete set differ from each other in the term
$\up^i\cR_i$, where $\up$ are vertical vector fields 
(\ref{cmp21}). If follows from the splitting (\ref{N20}) that this term
vanishes on the Lagrangian constraint space. The corresponding  constrained
Hamiltonian form
$H_N=i_N^*H$ and the  constrained Hamilton equations (\ref{N44}) can be
written. In the case of quadratic Lagrangians, we can improve Proposition
\ref{mm72} as follows.

\begin{prop} \label{cmp23}  For every Hamiltonian
form $H$ (\ref{N22}),
the Hamilton equations (\ref{m41b}) and (\ref{N28}) restricted to the
Lagrangian constraint space $N_L$  are equivalent to the constrained Hamilton
equations.
\end{prop}

\begin{proof} Due to the splitting (\ref{N20}), we have the corresponding 
splitting 
of the vertical tangent bundle $V_QV^*Q$ of the bundle $V^*Q\to Q$.
In particular, any
vertical vector field
$u$ on
$V^*Q\to \bR$ admits the decomposition
\be
u= [u-u_{TN}] + u_{TN},  \qquad  u_{TN}=u^i\dr_i +a_{ij}\si^{jk}_0u_k\dr^i, 
\ee
such that $u_N=u_{TN}\mid_{N_L}$ is a vertical vector field on the Lagrangian
constraint space $N_L\to \bR$. Let us consider the equations
\beq
r^*(u_{TN}\rfloor dH)=0 \label{cmp15}
\eeq
where $r$ is a section of $V^*Q\to \bR$ and $u$ is an arbitrary vertical vector
field on $V^*Q\to \bR$. They are equivalent to the pair of equations
\bea
&& r^*(a_{ij}\si^{jk}_0\dr^i\rfloor dH)=0,
\label{b4125a} \\
&& r^*(\dr_i\rfloor dH)=0. \label{b4125b}
\eea
The equations (\ref{b4125b}) are obviously the Hamilton equations
(\ref{m41b}) for $H$. Bearing in mind the relations (\ref{250}) and
(\ref{N21}), one can easily show that the equations (\ref{b4125a}) 
coincide with the Hamilton equations (\ref{N28}). The proof is
completed by observing that, restricted to the Lagrangian constraint space
$N_L$, the equations (\ref{cmp15}) are exactly the constrained Hamilton
equations (\ref{N44'}).
\end{proof} 

Proposition \ref{cmp23} shows that, restricted to the Lagrangian constraint
space, the Hamilton equations for different Hamiltonian forms (\ref{N22})
associated with the same quadratic Lagrangian (\ref{N12}) differ from each
other in the equations (\ref{N23}). These equations are independent of 
momenta and play the role of gauge-type conditions. 

We aim to obtain the Koszul--Tate resolution for the constraints
(\ref{zzz})  Since these constraints are not necessarily irreducible, we need 
an infinite number of ghosts and antighosts \cite{fisch,kimura}.

\section{Simple BRST manifolds}

Let $E=E_0\oplus E_1\to Z$ be the Whitney sum of vector bundles $E_0\to
Z$ and $E_1\to Z$ over a paracompact manifold $Z$. One can think of $E$ as
being a bundle of vector superspaces with a typical fibre $V=V_0\oplus V_1$
where transition functions of $E_0$ and $E_1$ are independent. Let us consider
the exterior bundle   
\beq
\w E^*=\op\bigoplus^\infty_{k=0} (\op\w^k_Z E^*), \
\label{mm80}
\eeq
which is the tensor bundle $\ot E^*$ modulo  elements 
\be
e_0e'_0 - e'_0e_0, \quad e_1e'_1 + e'_1e_1, \quad e_0e_1 - e_1e_0\quad
e_0,e'_0\in E_{0z}^*,
\quad e_1,e'_1\in E_{1z}^*, \quad z\in Z.
\ee 
$\w E^*$ is the bundle of commutative superalgebras $\w V$ which is the tensor
product $\vee E_0^*\ot\w E_1^*$ modulo elements 
\be
e_0e_1 - e_1e_0\quad
e_0\in E_{0z}^*,
\quad e_1\in E_{1z}^*, \quad z\in Z.
\ee
The global sections of $\w E^*$ constitute a commutative superalgebra $\cA(Z)$
over the free
$C^\infty(Z)$-module $E_0^*(Z)\oplus E_1^*(Z)$ of global sections of $E^*$.
This is the product of the commutative algebra $\cA_0(Z)$ of global sections of
$\vee E_0^*\to Z$ and the graded algebra $\cA_1(Z)$ of global sections of the
Grassman bundle
$\w E_1^*\to Z$.  We use the notation $\nw .$ for the Grassman parity.

\begin{rem}
Let $\cA_1$ be the sheaf of sections of the Grassman bundle $\w E_1^*$. The
pair $(Z,\cA_1)$ is a graded manifold \cite{bart}.
By the well-known Batchelor theorem, every graded manifold is isomorphic to a
sheaf of sections of some Grassman bundle, but not in a canonical way.
Therefore, the construction below can be extended to an arbitrary commutative
superalgebra over a free $C^\infty(Z)$-module $\cA=\cA_1\oplus\cA_2$ of finite
rank.  We call $(Z,\cA)$ a BRST manifold,  while sections of $\w E^*$ are
said to be BRST functions.
\end{rem} 

Let us study the $\cA(Z)$-module $\der \cA(Z)$ of graded derivations of
$\cA(Z)$. Recall that by a graded derivation of the commutative superalgebra
$\cA(Z)$ is meant an endomorphism of $\cA(Z)$ such that
\beq
 u(ff')=u(f)f'+(-1)^{\nw u\nw f}fu (f') \label{mm81}
\eeq
for the homogeneous elements $u\in \der\cA(Z)$ and $f,f'\in \cA(Z)$.

\begin{prop}
Graded derivations (\ref{mm81}) are represented by
sections of a vector bundle.
\end{prop}

\begin{proof}
Let $\{c^a\}$ be the holonomic bases for $E^*\to Z$ with respect to some bundle
atlas $(z^A,v^i)$ of $E\to Z$ with transition functions $\{\rho^a_b\}$, i.e.,
$c'^a=\rho^a_b(z)c^b$. Then BRST functions read
\beq
f=\op\sum_{k=0} \frac1{k!}f_{a_1\ldots
a_k}c^{a_1}\cdots c^{a_k}, \label{z785}
\eeq
where $f_{a_1\cdots
a_k}$ are local functions on $Z$, and we omit the symbol of an exterior product
of elements $c$. The coordinate transformation law of BRST functions
(\ref{z785}) is obvious. 
Due to the canonical splitting
$VE= E\times E$, the vertical tangent bundle 
$VE\to E$ can be provided with the fibre bases $\{\dr_a\}$ dual of $\{c^a\}$.
These are fibre bases for $\pr_2VE=E$. Then
any derivation $u$ of $\cA(U)$ on a trivialization domain $U$ of $E$ reads
\beq
u= u^A\dr_A + u^a\dr_a, \label{mm83}
\eeq
where $u^A, u^a$ are local BRST functions and $u$ acts on $f\in \cA(U)$ by
the rule
\beq
u(f_{a\ldots b}c^a\cdots c^b)=u^A\dr_A(f_{a\ldots b})c^a\cdots c^b +u^a
f_{a\ldots b}\dr_a\rfloor (c^a\cdots c^b). \label{cmp50'}
\eeq
This rule implies the corresponding
coordinate transformation law 
\beq
u'^A =u^A, \qquad u'^a=\rho^a_ju^j +u^A\dr_A(\rho^a_j)c^j \label{lmp2}
\eeq
of derivations (\ref{mm83}).
Let us consider 
the vector bundle
$\cV_E\to Z$ which is locally isomorphic to the vector bundle
\be
\cV_E\mid_U\approx\w E^*\op\ot_Z(\pr_2VE\op\oplus_Z TZ)\mid_U,
\ee
and has the transition functions
\be
&& z'^A_{i_1\ldots i_k}=\rho^{-1}{}_{i_1}^{a_1}\cdots
\rho^{-1}{}_{i_k}^{a_k} z^A_{a_1\ldots a_k}, \\
&& v'^i_{j_1\ldots j_k}=\rho^{-1}{}_{j_1}^{b_1}\cdots
\rho^{-1}{}_{j_k}^{b_k}\left[\rho^i_jv^j_{b_1\ldots b_k}+ \frac{k!}{(k-1)!} 
z^A_{b_1\ldots b_{k-1}}\dr_A(\rho^i_{b_k})\right] 
\ee
of the bundle coordinates $(z^A_{a_1\ldots a_k},v^i_{b_1\ldots b_k})$,
$k=0,\ldots$. These transition functions
fulfill the cocycle relations. It is readily observed that, for any
trivialization domain $U$, the
$\cA$-module $\der\cA(U)$ with the transition functions (\ref{lmp2}) is
isomorphic to the $\cA$-module of local sections of $\cV_E\mid_U\to U$.
One can show that, if $U'\subset U$ are open
sets, there is the restriction morphism $\der\cA(U)\to
\der\cA(U')$. It follows that, restricted to an open subset $U$, every
derivation $u$ of
$\cA(Z)$ coincides with some local section $u_U$ of $\cV_E\mid_U\to U$, whose
collection $\{u_U, U\subset Z\}$ defines uniquely a global section of
$\cV_E\to Z$, called a BRST vector field on $Z$. BRST vector
fields constitute  a Lie
superalgebra with respect to the bracket 
\be
[u,u']=uu' + (-1)^{\nw u\nw{u'}+1}u'u.
\ee
\end{proof}

\begin{cor}
The sheaf of sections of $\cV_E\to Z$ is isomorphic to the sheaf of graded
derivations of the sheaf $\cA$.
\end{cor}

There is the exact sequence over $Z$ of vector
bundles
\beq
0\to \w E^*\op\ot_Z\pr_2VE\to\cV_E\to \w E^*\op\ot_Z TZ\to 0. \label{cmp92}
\eeq
Its splitting 
\beq
\wt\g:\dot z^A\dr_A \mapsto \dot z^A(\dr_A +\wt\g_A^a\dr_a) \label{cmp70}
\eeq
 transforms every vector field $\tau$ on $Z$
into a BRST vector field 
\be
\tau=\tau^A\dr_A\mapsto \nabla_\tau=\tau^A(\dr_A +\wt\g_A^a\dr_a),
\ee
which is the derivation $\nabla_\tau$ of $\cA(Z)$ such that 
\be
\nabla_\tau(sf)=(\tau\rfloor ds)f +s\nabla_\tau(f), \quad f\in\cA(Z),\quad s\in
C^\infty(Z).
\ee
Thus, one can think of the splitting (\ref{cmp70}) as being a 
BRST connection on $Z$. 
For instance, every linear connection 
\be
\g=dz^A\ot (\dr_A +\g_A{}^a{}_bv^b \dr_a) 
\ee
on the vector bundle $E\to Z$ yields the BRST connection 
\beq
\g_S=dz^A\ot (\dr_A +\g_A{}^a{}_bc^b\dr_a) \label{cmp73}
\eeq
on $Z$ such that, for any vector field $\tau$ on $Z$ and any BRST function $f$,
the graded derivation $\nabla_\tau(f)$ is exactly the covariant derivative
of $f$ relative to the connection $\g$.

The $\w E^*$-dual $\cV^*_E$ of $\cV_E$ is a vector bundle over $Z$
which is locally isomorphic to the vector bundle
\be
\cV^*_E\mid_U\approx \w E^*\op\ot_Z(\pr_2VE^*\op\oplus_Z T^*Z)\mid_U,
\ee
and has the transition functions
\be
&& v'_{j_1\ldots j_kj}= \rho^{-1}{}_{j_1}^{a_1}\cdots
\rho^{-1}{}_{j_k}^{a_k} \rho^{-1}{}_j^a v_{a_1\ldots a_ka}, \nonumber\\
&& z'_{i_1\ldots i_kA}=
\rho^{-1}{}_{i_1}^{b_1}\cdots
\rho^{-1}{}_{i_k}^{b_k}\left[z_{b_1\ldots b_kA}+ \frac{k!}{(k-1)!} 
v_{b_1\ldots b_kj}\dr_A(\rho^j_{b_k})\right] 
\ee
of the bundle coordinates $(z_{a_1\ldots a_kA},v_{b_1\ldots b_kj})$,
$k=0,\ldots$, with respect to the dual bases $\{dz^A\}$ for $T^*Z$ and
$\{dc^b\}$ for $\pr_2V^*E=E^*$.
 Global sections of this vector bundle
constitute the $\cA(Z)$-module of  exterior BRST
1-forms $\f=\f_A dz^A + \f_adc^a$
on $Z$, which have the coordinate transformation law
\be
\f'_a=\rho^{-1}{}_a^b\f_b, \qquad \f'_A=\f_A
+\rho^{-1}{}_a^b\dr_A(\rho^a_j)\f_bc^j.
\ee
Then
the morphism $\f:u\to \cA(Z)$ can be seen as the interior product 
\beq
u\rfloor \f=u^A\f_A + (-1)^{\nw{\f_a}}u^a\f_a. \label{cmp65}
\eeq
There is the exact sequence
\beq
0\to \w E^*\op\ot_ZT^*Z\to\cV^*_E\to \w E^*\op\ot_Z \pr_2VE^*\to 0.
\label{cmp72}
\eeq
Any BRST connection $\wt\g$ (\ref{cmp70}) yields the
splitting of the exact sequence (\ref{cmp72}), and defines the corresponding
decomposition of BRST 1-forms
\be
\f=\f_A dz^A + \f_adc^a =(\f_A+\f_a\wt\g_A^a)dz^A +\f_a(dc^a
-\wt\g_A^adz^A). 
\ee

BRST $k$-forms $\f$ can be defined as sections
of the graded exterior bundle $\ol\w^k_Z\cV^*_E$ such that
\be
 \f\ol\w\si =(-1)^{\nm\f\nm\si +\nw\f\nw\si}\si\ol\w \f.  
\ee
The interior product (\ref{cmp65})
is extended to higher BRST forms by the rule  
\be
u\rfloor (\f\ol\w\si)=(u\rfloor \f)\ol\w \si
+(-1)^{\nm\f+\nw\f\nw{u}}\f\ol\w(u\rfloor\si). 
\ee
The graded exterior differential
$d$ of BRST functions is introduced by the condition 
$u\rfloor df=u(f)$
for an arbitrary BRST vector field $u$, and  is
extended uniquely to higher BRST forms by the rules
\be
d(\f\ol\w\si)= (d\f)\ol\w\si +(-1)^{\nm\f}\f\ol\w(d\si), \qquad  d\circ d=0.
\ee
It takes the coordinate form
\be
d\f= dz^A \ol\w \dr_A(\f) +dc^a\ol\w \dr_a(\f), 
\ee
where the left derivatives 
$\dr_A$, $\dr_a$ act on the coefficients of BRST forms by the rule
(\ref{cmp50'}), and they are graded commutative with the forms $dz^A$, $dc^a$.
The Lie
derivative of a BRST form $\f$ along a BRST vector field $u$ is given by
the familiar formula
\be
\bL_u\f= u\rfloor d\f + d(u\rfloor\f). 
\ee

\section{The Koszul--Tate resolution}

To construct the vector bundle $E$ of antighosts, let
us consider the vertical tangent bundle $V_Q(V^*Q)$ of $V^*Q\to Q$.
Let us chose the bundle $E$ as the Whitney sum of the
bundles $E_0\oplus E_1$ over $V^*Q$ which are the infinite Whitney sum
over
$V^*Q$ of the copies of
$V_Q(V^*Q)$.  We have
\beq
E= V_Q(V^*Q)\op\oplus_{V^*Q}V_Q(V^*Q)\oplus\cdots.
\label{mm84}
\eeq
This bundle is provided with the holonomic coordinates $(t,q^i,p_i,\dot
p_i^{(k)})$, $k=0,1,\ldots$, where  $(t,q^i,p_i,\dot
p_i^{(2r)})$ are coordinates on $E_0$, while 
$(t,q^i,p_i,\dot p_i^{(2r+1)})$ are those on $E_1$. We call
$k$ the antighost number, while $k\,{\rm mod2}$ is the Grassman parity. The
dual of
$E\to V^*Q$ is
\be
E^*= V^*_Q(V^*Q)\op\oplus_{V^*Q}V^*_Q(V^*Q)\oplus\cdots.
\ee
It is  endowed with the
associated fibre bases
$\{c_i^{(k)}\}$, $k=1,2,\ldots$, such that
$c_i^{(k)}$ have the same linear coordinate
transformation law as the coordinates $p_i$. The corresponding
BRST vector fields and BRST forms are introduced on $V^*Q$ as sections of the
vector bundles $\cV_E$ and $\cV^*_E$, respectively. 

The $C^\infty(V^*Q)$-module $\cA(V^*Q)$ of BRST functions is graded by the
antighost number as 
\be
\cA(V^*Q)=\op\oplus_{k=0}^\infty \cN^k, \qquad \cN^0=C^\infty(V^*Q).
\ee
Its terms $\cN^k$ constitute a complex
\beq
0\lla C^\infty(V^*Q) \lla \cN^1\lla \cdots \label{mm90}
\eeq
with respect to the Koszul--Tate  differential
\ben
&& \dl: C^\infty(V^*Q)\to 0, \nonumber \\
&& \dl(c^{2r}_i)= a_{ij}\si^{jk}c^{2r-1}_k, \qquad r>0, \label{mm91}\\
&& \dl(c^{2r+1}_i)=(\dl_i^k- a_{ij}\si^{jk})c^{2r}_k, \qquad, r>0, \nonumber\\
&&  \dl(c^1_i)=(\dl_i^k- a_{ij}\si^{jk})p_k. \nonumber
\een
The nilpotency property $\dl\circ\dl=0$ of this differential is the corollary
of the relations (\ref{mm100}) and (\ref{N21}).

\begin{prop}
The complex (\ref{mm90}) with respect to the differential (\ref{mm91}) is the
Koszul--Tate resolution, i.e., its homology groups are
\be
H_{k>1}=0, \qquad H_0=C^\infty(V^*Q)/I_N=C^\infty(N_L). 
\ee
\end{prop}

Note that, in particular cases of the degenerate quadratic Lagrangian
(\ref{N12}), the complex (\ref{mm90}) may have a subcomplex, which is also
the Koszul--Tate resolution. For instance, if the fibre metric $a$ in $VQ\to
Q$ is diagonal with respect to a holonomic atlas of $VQ$, the constraints
(\ref{zzz}) are irreducible and the complex (\ref{mm90}) contains a
subcomplex which consists only of the antighosts
$c_i^{(1)}$.

Now let us construct the BRST
charge $\bQ$ such that 
\be
\dl(f)=\{\bQ, f\}, \qquad f\in \cA(V^*Q)
\ee
with respect to some Poisson bracket. The problem is to find the Poisson
bracket such that $\{f,g\}=0$ for all $f,g\in C^\infty(V^*Q)$. To overcome
this difficulty, one can consider the vertical extension of Hamiltonian
formalism onto the configuration bundle $VQ\to\bR$ \cite{giach99,book98}.

\begin{lem} \label{mm81'} 
Given a fibre bundle $Y\to X$, there is the isomorphism 
\be
VV^*Y \op\cong_{VY} V^*VY, \quad p_i\llra\dot v_i, 
\quad \dot p_i\llra\dot y_i. 
\ee
\end{lem}

\begin{proof}
The proof is based on inspection of the
transformation laws of the holonomic
coordinates $(x^\la, y^i, p_i)$ on
$V^*Y$ and $(x^\la, y^i, v^i)$ on $VY$. 
\end{proof} 

Given a configuration bundle $Q\to\bR$, let us consider the vertical
tangent bundle $VQ\to\bR$, seen as a configuration bundle of the above
mentioned vertical extension of Hamiltonian formalism. By
virtue of Lemma
\ref{mm81'}, the corresponding Legendre bundle $V^*(VQ)$  is isomorphic to
$V(V^*Q)$, and is provided with the holonomic coordinates $(t,q^i,p_i,\dot
q^i,\dot p_i)$ such that $(q^i,\dot p_i)$ and $(\dot q^i, p_i)$ are 
conjugate pairs of canonical coordinates. The momentum phase space $V(V^*Q)$ is
endowed with the canonical exterior 3-form
\beq
\bom_V=\dr_V\bom=[d\dot p_i\w dq^i +dp_i\w d\dot q^i]\w dt,
\label{m145}
\eeq
where we use the compact notation
\be
\dot\dr_i=\frac{\dr}{\dr\dot q^i}, \quad \dot\dr^i=\frac{\dr}{\dr\dot
p_i}, \quad \dr_V=\dot q^i\dr_i +\dot p_i\dr^i.
\ee
The corresponding Poisson bracket on $V(V^*Q)$ reads
\be
\{f,g\}_{VV} =\dot\dr^if\dr_ig +\dr^if\dot\dr_ig -\dr^ig\dot\dr_if
-\dot\dr^ig\dr_if. 
\ee

To extend this bracket to BRST functions, let us consider the following
graded extension of Hamiltonian formalism \cite{gozz,book98}.
We will assume that $Q\to\bR$ is a vector bundle, and will further denote
$\Pi=V^*Q$.

Let us consider the vertical tangent bundle $VV\Pi$. It
admits the canonical decomposition 
\beq
VV\Pi=V\Pi\op\oplus_\bR V\Pi\ar^{\pr_1} V\Pi. \label{cmp68}
\eeq
Let choose the bundle $E$ as the
Whitney sum of the bundles $E_0\oplus E_1$ over $V\Pi$ which are the
infinite Whitney sum over
$V\Pi$ of the copies of
$VV\Pi$. In view of the decomposition (\ref{cmp68}), we have
\be
E=V\Pi\op\oplus_\bR V\Pi\oplus\cdots\op\to^{\pr_1} V\Pi.
\ee
This bundle is provided with the holonomic coordinates $(t,q^i,p_i,\dot
q^i_{(k)},\dot p_i^{(k)})$, $k=0,1,\ldots$, where  $(t,q^i,p_i,\dot
q^i_{(2r)},\dot p_i^{(2r)})$ are coordinates on $E_0$ and 
$(t,q^i,p_i,\dot
q^i_{(2r+1)},\dot p_i^{(2r+1)})$ are those on $E_1$.  The dual
of
$E\to V\Pi$ is
\be
E^*=V\Pi\op\oplus_\bR V\Pi^*
\oplus\cdots.
\ee
It is  endowed with the
associated fibre bases
$\{\ol c^i_{(k)},\ol c_i^{(k)},c^i_{(k)},c_i^{(k)}\}$, $k=1,\ldots$. The
corresponding BRST vector fields and BRST forms are introduced on $V\Pi$ as
sections of the vector bundles $\cV_E$ and $\cV^*_E$, respectively. Let us
complexify these bundles as $\bC\op\ot_\bR\cV_{VV\Pi}$ and
$\bC\op\ot_\bR\cV^*_{VV\Pi}$.  

The BRST extension of the form (\ref{m145}) on $V^*Q$ is the 3-form 
\beq
\Om_S=[d\dot p_i\w dq^i +dp_i\w d\dot q^i +i\op\sum_{k=1}^\infty(d\ol
c_i^{(k)}\w dc^i_{(k)}- dc_i^{(k)}\w d\ol c^i_{(k)})]\w dt 
\eeq
The corresponding bracket of BRST functions on $V^*Q$ reads
\ben
&&\{f,g\}_S=\{f,g\}_{VV} +i\op\sum_{k=1}^\infty (-1)^{k[f]}[
\frac{\dr f}{\dr \ol c_i^{(k)}}\frac{\dr g}{\dr c^i_{(k)}} + (-1)^k
\frac{\dr f}{\dr \ol c^i_{(k)}}\frac{\dr g}{\dr c_i^{(k)}}- \label{mm103}\\
&&\qquad \frac{\dr f}{\dr  c_i^{(k)}}\frac{\dr g}{\dr \ol c^i_{(k)}} - (-1)^k
\frac{\dr f}{\dr  c^i_{(k)}}\frac{\dr g}{\dr \ol c_i^{(k)}}]. \nonumber
\een
It satisfies the condition $\{f,g\}_S=-(-1)^{[f][g]}\{g,f\}_S$.
Then the desired  BRST charge takes the form
\be
\bQ=i[\ol c^i_{(1)}(\dl_i^k - a_{ij}\si^{jk})p_k + \op\sum_{r=1}^\infty (\ol
c^i_{(2r)}a_{ij}\si^{jk}c_k^{(2r-1)} + \ol
c^i_{(2r+1)}(\dl_i^k-a_{ij}\si^{jk})c_k^{(2r)})].
\ee
Using the bracket (\ref{mm103}), one can extend this charge in order to obtain
the BRST complex for antighosts
$c_i^{(k)}$ and ghosts $\ol c^i_{(k)}$.

\end{document}